# A Framework for Modeling Liquefaction-Induced Road Disruptions After Earthquakes: Implications for Emergency Response and Access in the Cascadia Region of North America

Morgan D. Sanger[1], Olyvia B. Smith[2], Brett W. Maurer[3],
Liam Wotherspoon[4], Marc O. Eberhard[5], and Jeffrey W. Berman[6]

**ABSTRACT**

Large earthquakes along the Cascadia Subduction Zone (CSZ) are expected to trigger widespread soil liquefaction that could disrupt transportation systems across the U.S. Pacific Northwest. However, past regional assessments have relied on simple geologic screening methods and binomial shaking thresholds that are only loosely informed by liquefaction science. This study introduces a mechanics-informed, data-driven framework for estimating liquefaction-induced road closures and service reductions, and the framework is applied to a magnitude-9 CSZ earthquake. Liquefaction hazard is mapped using a geospatial liquefaction model trained on more than 37,000 cone penetration tests and conditioned on regional datasets describing geomorphology, geology, climate, and hydrology. Predicted liquefaction severity is translated into segment-level probabilities of closure and reduced service using empirically derived fragility relationships. These probabilities are mapped at 90-m resolution across Washington, Oregon, and California, and propagated through the National Highway System using a spatially correlated Monte Carlo simulation to estimate link-level disruption. Results show that impacts are concentrated in low-lying coastal zones, river valleys, and urban waterfronts, with major disruptions expected along critical routes including U.S. Route 101. Local mobility is further examined in Pacific and Grays Harbor counties, Washington,

[1] Dept. of Civil and Environmental Engineering, University of Washington, Seattle WA, USA, sangermd@uw.edu
[2] Dept. of Civil and Environmental Engineering, University of Washington, Seattle WA, USA, smitho2@uw.edu
[3] Dept. of Civil and Environmental Engineering, University of Washington, Seattle WA, USA, bwmaurer@uw.edu
[4] Dept. of Civil and Environmental Engineering, University of Auckland, Auckland, NZ, l.wotherspoon@auckland.ac.nz
[5] Dept. of Civil and Environmental Engineering, University of Washington, Seattle WA, USA, eberhard@uw.edu
[6] Dept. of Civil and Environmental Engineering, University of Washington, Seattle WA, USA, jwberman@uw.edu

where limited network redundancy, strong shaking, and high liquefaction susceptibility lead to elevated probabilities of isolation and loss of hospital access. Socioeconomic analysis reveals modest but statistically significant associations between road impacts and demographic indicators, suggesting that liquefaction impacts may compound with existing social vulnerabilities. While not a substitute for site-specific analysis, the results provide a regional baseline for emergency planning, risk communication, and prioritization of more advanced geotechnical sampling and analysis. Moreover, the methodology proposed here is not specific to the CSZ, but rather, could be applied to analogous studies of road impacts elsewhere.

**INTRODUCTION**

The Cascadia Subduction Zone (CSZ) is a 1,300-km-long convergent plate boundary along the Pacific coast of North America, extending from Cape Mendocino in northern California to Vancouver Island in southern British Columbia (e.g., Walton et al., 2021). Geological and historical evidence indicate the most recent megathrust rupture of the CSZ occurred in 1700 CE, as inferred from Japanese tsunami records and regional paleoseismic data (e.g., Atwater, 1987; Yamaguchi et al., 1997; Satake et al., 1996; Peters et al., 2007; Goldfinger et al., 2012; Obermeier, 1995; Atwater et al., 2014; Rasanen et al., 2021). This earthquake is believed to be one of at least 19 similar events over the past 10,000 years, each judged to be magnitude-8 (M8) to M9 or greater (e.g., Walton et al., 2021). Although there are few oral histories of any prior CSZ rupture (Thrush and Ludwin, 2007), it is expected that the next could cause devastating effects across the region, as has been vividly described in popular media (e.g., Schulz, 2015).

Soil liquefaction – defined as the loss of strength and stiffness in saturated soils due to shaking-induced pore pressure – could be among the most widespread and economically consequential effects, as demonstrated by the 2010-2011 Canterbury, New Zealand, earthquakes, where liquefaction caused an estimated $10B in damage (Parker and Steenkamp, 2012). Liquefaction can damage a variety of infrastructure (e.g., buildings, pipelines, levees, pavements, rail lines, and bridges) and, due to its impacts on transportation assets, also hinder post-event emergency response, recovery, and access to critical facilities (e.g., hospitals, fire stations, food distribution centers, transportation hubs, schools). In the case of road networks, liquefaction can lead to large deformation in pavements, damage to bridge approaches,

abutments, and foundations, and inundation by liquefaction ejecta, obscuring road geometry and impeding travel consistent with snowfall or flooding. The cumulative result is a degradation in functionality (i.e., reductions in travel speeds and road capacity) and potential closure (e.g., Kajihara et al., 2025).

The U.S. Department of Homeland Security (DHS), for example, studied ~15,000 km of road in western Washington (WA) and predicted that in the wake of an M9 CSZ rupture, a total of 2,755 km, or 18%, would be unavailable due to liquefaction (DHS, 2019). Notably, these analyses considered only state-owned roads. DHS (2019) further predicted that of 6,911 bridges in western WA, 2,669, or 39%, will sustain "significant" damage due to liquefaction and be unusable, with an average repair time of 462 days to a minimum acceptable state for emergency response and supply vehicles. Although these analyses adopted many simplifying assumptions, as soon discussed, they emphasize the potentially staggering impact of liquefaction on post-event mobility in CSZ ruptures.

Liquefaction models grounded in subsurface geotechnical measurements, such as those from the cone penetration test (CPT), were first introduced in the 1970s (e.g., Seed & Idriss, 1971). Since then, larger datasets and advances in liquefaction science have produced a suite of models that are widely used in engineering practice, with global prediction efficiencies typically in the range of 75-85% (e.g., Geyin et al., 2020). An ensemble of popular models was applied to 24 case-histories from the 2001 Nisqually, WA, earthquake, for instance, and achieved prediction efficiencies of ~90% (Rasanen et al., 2023). Using these models, liquefaction in CSZ events has been predicted at some discrete sites (e.g., Rasanen et al., 2025; Rasanen, 2025), but because subsurface geotechnical data cannot be collected continuously across large areas, it is inherently challenging to predict liquefaction at regional scale. As a result, many such predictions have been resigned to infer soil traits from geologic maps. The DHS (2019) predictions, for example, adopted a "HAZUS" (FEMA, 2017) style of approach in which broad geologic units were assigned susceptibility classes of "low," "moderate," or "high," with liquefaction expected when peak ground accelerations ($PGA$s) exceeded class-specific thresholds. These thresholds ranged from 0.1 g for "high" susceptibility units to 0.15 g for those with "low" susceptibility. If the $PGA$ exceeded this threshold, liquefaction was assumed to occur, and any coincident bridge was assumed to suffer "significant" damage due to liquefaction, resulting in closure. While these predictions provide a semblance of where liquefaction could occur, they are unaware of local subsurface conditions and adhere only loosely to liquefaction science. We believe the outcomes – which in WA alone include closure of over 2,500 bridges and 2,500-km of state-owned roadway – are overly conservative and might induce fatalistic perspectives. It is unclear how these coarse predictions could be used by decision-makers to prioritize retrofits, direct more advanced geotechnical sampling and analysis, or plan for future earthquakes.

We aim to predict impacts to roads in CSZ ruptures more accurately using the emergent geospatial liquefaction model (GLM) of Sanger et al. (2025), a mechanics-informed machine learning (ML) framework trained on over 37,000 CPTs. The GLM outputs values of the liquefaction potential index (*LPI*), a widely used predictor of ground failure and damage potential (Iwasaki, 1978), by estimating subsurface traits via dozens of geospatial correlates. This model has conceptual benefits over prior regional-scale methods in that its predictions: (i) are anchored to geotechnical models and the knowledge of liquefaction mechanics embedded therein; (ii) use ML to exploit a large library of predictive information; and (iii) are geostatistically updated near subsurface measurements to reflect measured conditions. This model also showed statistically significant improvement in global testing (Sanger et al., 2025).

By overlaying GLM predictions with road networks in WA, Oregon (OR), and California (CA), we probabilistically predict reduced service and closure at both the local segment-level and broader link-level, thus quantifying impacts across various scales and populations. We also model accessibility to critical facilities in two WA counties of special interest: Pacific and Grays Harbor, where shaking intensities are high, liquefaction-susceptible soils are common, and road-networks often lack redundancy. Additionally, we predict liquefaction impacts to WA bridges and contrast our results against DHS (2019). Our aims and methods are partly inspired by Lin et al. (2024), who evaluated exposure of New Zealand roads to liquefaction using an earlier GLM. Our results contribute to seismic resilience in the region by supporting emergency response and recovery planning, and by improving prioritization of assets for focused analysis.

## DATA AND METHODOLOGY

### Road Networks

To evaluate impacts to roadways most crucial to the region's economy, welfare, and mass mobility, we adopt the Federal Highway Administration (2025) "National Highway System" (NHS) dataset, which consists of interstate highways and state roads, as well as other principal arterials that either provide access to ports, airports, or other transportation facilities; connect defense facilities and major highways; or bridge any of the subsystems making up the NHS. We augment this dataset with a small number of state-owned roads missing from the NHS using datasets in WA (Washington State Department of Transportation, 2024), OR (Oregon Department of Transportation, n.d.), and CA (California Department of Transportation, 2025). The resulting composite road network is mapped in Fig. 1 for the study area. The two counties of special interest (Pacific and Grays Harbor) are delineated for further analyses, wherein we adopt the extensive road network of OpenStreetMap (2025) excluding sidewalks, bike paths, trails, bridleways, unpaved forest roads, and "service roads," which are generally parking lots, alleyways, or driveways used to access residences and businesses. The resulting local network is shown later. All roads were spatially discretized at a resolution of 0.000833° (~90 m) to match that of the Sanger et al. (2025) GLM. For the NHS, this resulted in ~509,000 discretized road sections, which we refer to as "segments," within the extent of the U.S. mapped in Fig. 1. These segments were then grouped between network nodes (i.e., intersections of two or more NHS roads) and assigned to a parent, which we refer to as "links." The resulting ~12,000 links reflect the expected performance of many segments in series and illustrate where travel routes lack redundancy. Links thus facilitate an understanding of how segment-level impacts propagate to network-level performance.

**CSZ Earthquake Scenarios**

To model a CSZ event, we adopt the median ground motions from the Wirth et al. (2021) catalog of CSZ M9 scenario ruptures. These estimates of shaking were derived from 30 physics-based realizations with differing hypocenter locations, down-dip rupture limits, slip distributions, and locations of strong-motion-generating subevents (Frankel et al., 2018). Whereas traditional scenario studies are typically based on a single hypothetical rupture, Wirth et al. (2021) used a logic-tree approach to ensemble motions from multiple scenarios. Their simulations may also better represent salient phenomena relative to empirical models, such as amplification in the region's sedimentary basins at longer periods. In this regard, Wirth et al. (2021) adopted a region-specific ground velocity model and propagated the simulated motions through this velocity structure using wave-propagation-based site response analyses. The estimated median *PGA*s across the region are shown in Fig. 1. These *PGA*s typically exceed 0.5 g along the coast and decay to < 0.1 g approximately 175 km inland, aligning with the north-south Cascade Mountain range.

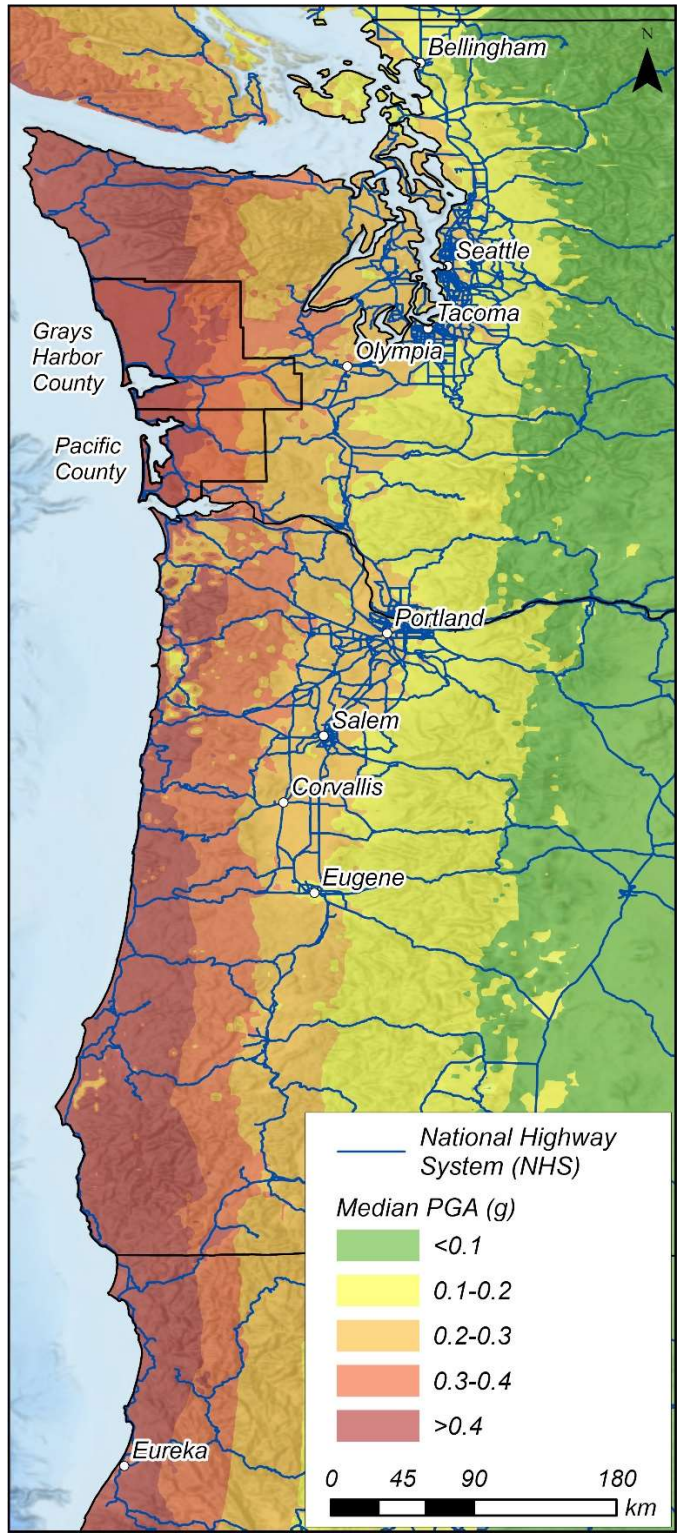

**Figure 1.** Predicted median *PGA*s in an M9 CSZ rupture.

**Geospatial Liquefaction Model (GLM)**

The GLM of Sanger et al. (2025) uses geospatial data and ML to surrogate, or mimic, CPT-based liquefaction models and was trained at the sites of 37,000 global CPTs. This mechanistic anchorage encourages more sensible scaling and response across the domains of soil, site, and earthquake loading conditions than could be achieved by a purely data-driven model. To predict values of *LPI* without in-situ data, the GLM uses 37 geospatial predictors that correlate to subsurface conditions relevant to liquefaction susceptibility and response, including metrics of topography, hydrology, geomorphology, geology, and climate. Predictions are then geostatistically updated near CPT measurements, ensuring that results are consistent with subsurface conditions, where sampled. Formally, the *LPI* at a given location is expressed as a function of ML-predicted parameters (*A*, *B*) and the magnitude-scaled *PGA* (*PGA$_M$*) (Eq. 1):

$$LPI = \begin{cases} 0, for\ PGA_M < \frac{A}{100B}g \\ A*(\tan^{-1}(B*(PGA_M - \frac{A/100}{B})^2)), for\ PGA_M \geq \frac{A}{100B}g \end{cases} \quad (Eq.\ 1)$$

where *A* and *B* are fitting parameters that allow ML predictions to track the response predicted by a geotechnical model, tan$^{-1}$ is expressed in radians, and *g* is the gravitational constant. *LPI* is a state-of-practice manifestation model that estimates liquefaction consequences at the ground surface using a depth-weighted integration of the computed factor of safety against liquefaction triggering (Iwasaki, 1978), and is further discussed in the *Supplemental Materials* (Eq. S1). For detailed coverage of the GLM development and performance, the reader is directed to Sanger et al. (2025).

Fig. 2 maps *LPI* predicted by the GLM for the median *PGA*s expected in an M9 CSZ rupture, highlighting both regional patterns and some cities of interest. Across the study area (Fig. 2a), the highest *LPI* values are found where the expected shaking is severe and where susceptible soils are common, such as low-lying coastal zones, river valleys, and estuarine environments. Within WA, localized hotspots of high *LPI* are evident in Seattle (Fig. 2b) and Olympia (Fig. 2c), particularly in waterfronts underlain by young alluvium and artificial fill. Grays Harbor (Fig. 2d) also exhibits broad zones of elevated hazard, especially on the peninsulas that underlay Ocean Shores and Westport. Southward, Portland, OR (Fig. 2e) shows high predicted *LPI* along the Columbia River. In Eureka, CA (Fig. 2f), high *LPI* values trace the margins of Humboldt Bay. Also mapped in Fig. 2 are the sites of CPTs used by Sanger et al. (2025) to train and locally update GLM predictions; these illustrate where predictions are informed by subsurface data and geotechnical models (generally urban areas), and where they are purely ML-driven (generally rural areas).

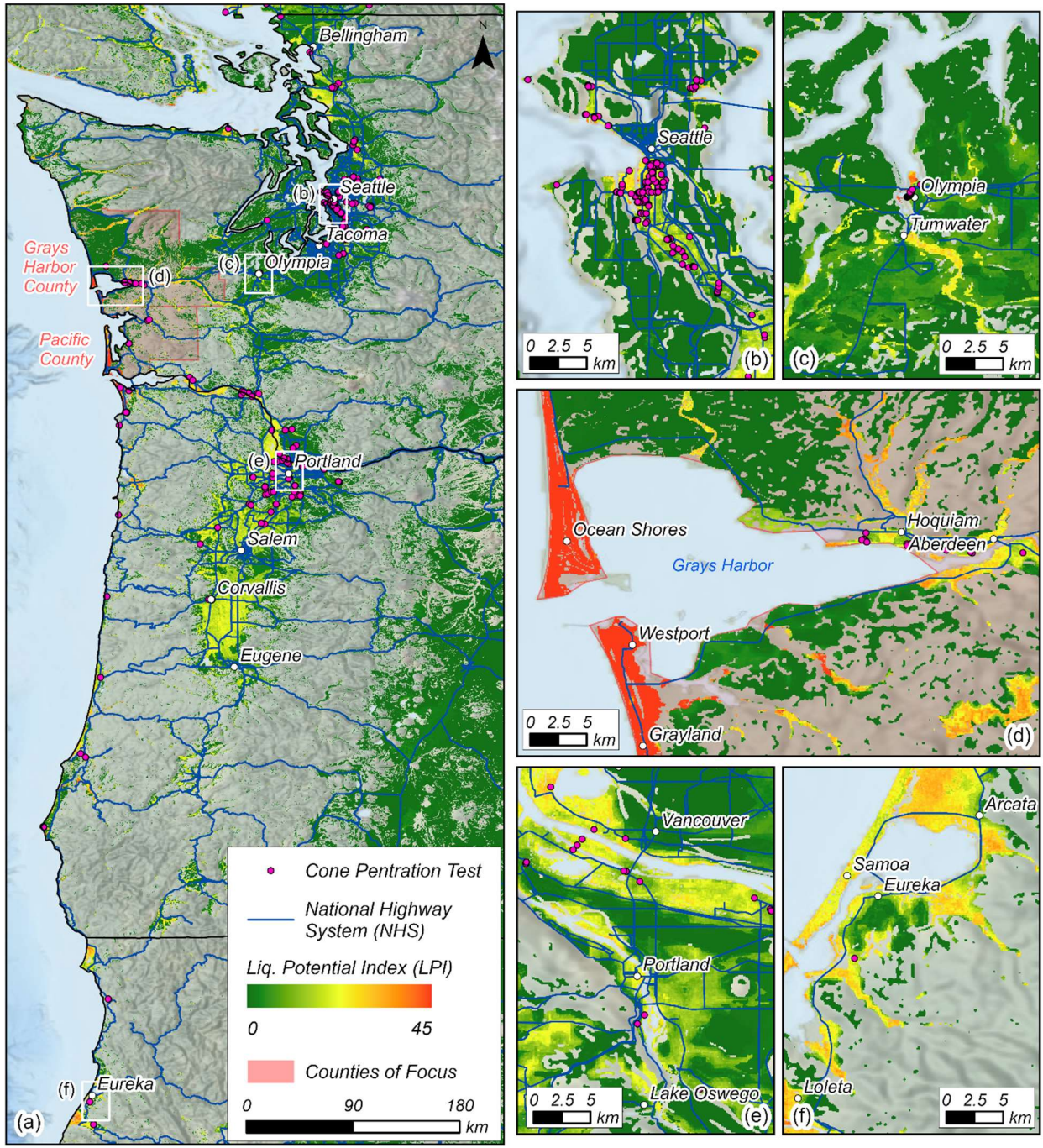

**Figure 2.** Predicted liquefaction potential index (*LPI*) in a median M9 CSZ rupture: (a) full study area; (b) Seattle, WA; (c) Olympia, WA; (d) Grays Harbor, WA; (e) Portland, OR; and (f) Eureka, CA.

**Predicting Segment-Level Closures and Service Reductions**

Enumerable studies have addressed liquefaction occurrence and consequences, but few, if any, have explicitly compiled data or trained models to estimate road disruptions. DHS (2019), for example, assumed that roads would be closed if any liquefaction occurred, which we view as overly conservative. In this study, we employ the fragility functions of Geyin and Maurer (2020), which are conditioned on *LPI* and predict the severity of surficial liquefaction manifestation. Given a predicted *LPI*, these functions forecast the probabilities of the damage state (*DS*) being (i) "none;" (ii) "minor," where roads have isolated or trace amounts of ejecta less than a vehicle width; (iii) "moderate," where 5% to 40% of the road surface is covered by ejecta; or (iv) "severe," where more than 40% of the ground surface is covered by ejecta and contiguous masses of liquefaction ejecta exceed a vehicle width. Shown in Fig. 3(a) are the probabilities of reaching or exceeding these states. At *LPI* = 5, for example, the probabilities of manifestations reaching or exceeding minor, moderate, and severe are 54%, 23%, and 2%, respectively. The functions can instead be expressed as the probabilities of being *in* each state, as shown in Fig. 3(b). Again, using *LPI* = 5 as an example, the probabilities of manifestations being none, minor, moderate, and severe are 46%, 31%, 21%, and 2%, respectively. We pair these predictions, which were trained on ~15,000 case histories, with studies on road disruptions from flooding, snow, and ashfall, to inform predictions of impacts from liquefaction.

Research on volcanic ashfall indicates that just 1 mm of broad coverage can greatly reduce service by obscuring road markings and reducing traction; at 10 cm depth, roads become impassable for some vehicles, and widespread impassability may occur around 20-30 cm (Blake et al., 2017). Observations of flooded and snow-covered roads suggest similar thresholds for closure and reduced service (e.g., Kramer et al., 2016; Umeda et al., 2021). Of course, any such threshold is uncertain and likely depends on the road surface and grade, vehicle clearance, and driver. Yet in the absence of studies specific to liquefaction, these thresholds and the case studies analyzed by Geyin and Maurer (2020) support reasonable estimates. Thus, we equate the probability of reduced service to that of "moderate" or greater liquefaction manifestation per Geyin and Maurer (2020); this assumes travel is unaffected by "minor" manifestations but that broader coverage by ejecta reduces service. We compute the closure probability of a road segment as (Eq. 2):

$$P(Closure|LPI) = \sum_i P(Closure|DS_i) \cdot P(DS_i|LPI) \quad (Eq.\ 2)$$

where $LPI$ is computed by the Sanger et al. (2025) model; $P(DS_i|LPI)$ is the probability of $DS_i$, as computed by Geyin and Maurer (2020) and plotted in Fig. 3(b); $i$ is the $DS$ index, which ranges from 1 (no manifestation) to 3 (severe manifestation); and $P(Closure|DS_i)$ is the probability of closure conditioned on $DS_i$, which for $i$ = 0, 1, 2, and 3 we respectively assign as 0, 0.01, 0.05, and 0.65. This means, for example, that there is a 65% chance of closure when liquefaction manifestations are severe, which is consistent with the Geyin and Maurer (2020) observation that roads were typically impassable by the modal vehicle. Of course, other analysts might reasonably justify other conditional probabilities. Integrating over all $DS_i$, $P(Closure|LPI)$ is computed and plotted in Fig. 3(a). This probability remains relatively low even at very high $LPI$ (e.g., 15.5% at $LPI$ = 30).

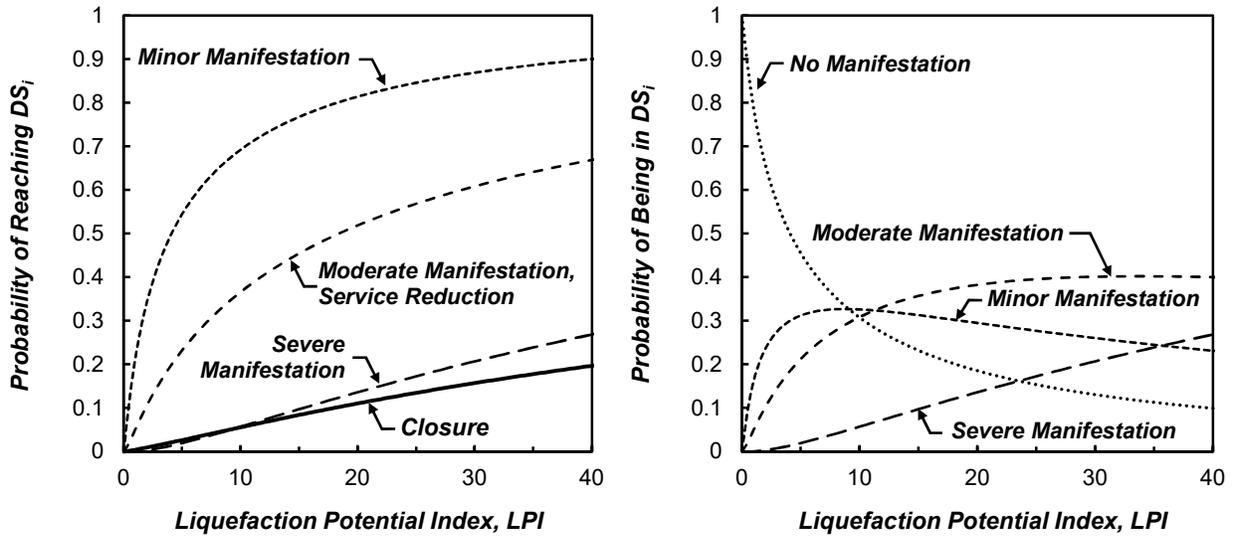

**Figure 3.** Probability of liquefaction manifestations: (a) reaching or exceeding a damage state, $DS_i$; and (b) being in a damage state, $DS_i$, conditioned on $LPI$ (Geyin and Maurer, 2020).

While limitations and uncertainties are discussed in a dedicated section, we note here that lateral spreading, a distinct expression of liquefaction, is not explicitly predicted by the Geyin and Maurer (2020) functions. This is because lateral spreading depends on topographic features that are not accounted for in the $LPI$ formula (e.g., distance to a free face). As a result, lateral spreading – which can cause more damage than ejecta, ground settlement, and cracking alone – can occur at $LPI$ values for which the expected damage severity would otherwise be less (e.g., Maurer et al., 2015). Although this remains a limitation and source of uncertainty, we attempt to address it by adopting non-zero conditional closure probabilities for "minor" (0.01) and "moderate" (0.05) manifestations, acknowledging that although these manifestations, as defined by Geyin and Maurer (2020), are not expected to cause closure, there is a non-zero probability of lateral spreading that could cause closure, and this probability increases with increasing *LPI*.

Following this approach, the probabilities of closure and service reduction are estimated for each 90-m segment and are input to four other analyses. First, the predictions are upscaled to link-level outcomes to contextualize impacts in terms of NHS connectivity. Second, to explore whether disruptions stratify socioeconomically, we perform a correlation analysis between segment closures and demographic metrics. Third, we interrogate the DHS (2019) predictions for 6,911 bridges in western WA – of which 39% were expected to be closed due to liquefaction – to provide estimates that, while still simplified, are argued to be more accurate. Fourth, post-event point-to-point mobility is modeled within the two WA counties of focus. Methods for the first and fourth of these analyses follow, while others are introduced in the results section.

**Predicting Link-Level Performance**

We compute both the probability that a link has at least one closure (Eq. 3) and the probability that a link has at least one service reduction (Eq. 4):

$$P\ (link\ closed) = 1 - \prod_{i=1}^{n}(1 - c_i) \quad \text{(Eq. 3)}$$

$$P\ (link\ service\ reduction) = 1 - \prod_{i=1}^{n}(1 - r_i) \quad \text{(Eq. 4)}$$

where $n$ is the number of segments in a link, $c_i$ is the closure probability for segment $i$, and $r_i$ is the service reduction probability for segment $i$. Because link-level outcomes depend on the joint behavior of many road segments, the treatment of spatial correlation is an important factor. Many links contain thousands of segments and while the likelihood of impacts to any one segment is typically very low (especially for segment closure), it is often highly likely that at least one segment per link is impacted if segments are treated as statistically independent. However, proximal road segments tend to have similar soil conditions, groundwater depths, and shaking intensities; thus, liquefaction impacts on roads are spatially correlated.

We examine correlation by computing semivariograms for logit-transformed probabilities, which is used to stabilize variance across the probability scale. A semivariogram describes how similarity between segment impacts decays with distance and is generically expressed as (Eq. 5):

$$\gamma(h) = \frac{1}{2} Var[Z(x) - Z(x+h)] \quad \text{(Eq. 5)}$$

where $\gamma(h)$ is the semivariance; $h$ is the lag distance, representing the distance between road segments; $Var$ is the variance, which measures the spread of the data; and $Z(x)$ and $Z(x+h)$ are the segment probabilities of link closure or service disruption at two different locations, separated by distance $h$. To fit our spatial correlation empirically, we adopt an exponential semivariogram (e.g., Cressie, 2015) for its best fit of the data (Eq. 6):

$$\gamma(h) = c_o \left(1 - \exp\left(-h/L_c\right)\right) \quad \text{(Eq. 6)}$$

where $c_o$ is the sill, controlling the maximum semivariance; $h$ is as previously defined; and $L_c$ is the correlation length, representing the distance at which correlation decays to approximately 1/e. Fitting this model to the segment closure probabilities gave $L_c$ = 1.68 km and $c_o$ = 0.71, indicating that the expected impacts to adjacent road segments are moderately correlated and that that correlation drops substantially beyond a few kilometers, reflecting the substantial subsurface variability that often exists over such scales. It should be noted that this correlation structure captures a regional average and is not a site-specific truth (e.g., it may not accurately translate across geologic transitions or engineered variability in urban settings).

To generate realistic realizations of road outcomes, we use the fitted correlation structure to generate correlated random fields via a Gaussian copula approach (e.g., Nelsen, 2006). This approach is summarized in three steps: (i) we create many simulated earthquakes in a way that respects both the segment-level probabilities and the correlation structure; (ii) for each simulation, we map segment results to a Bernoulli outcome (i.e., closed or not, service reduced or not) and check whether at least one segment is impacted in each link; and (iii) Monte Carlo simulation repeats the process 5,000 times for each of 12,305 road links, yielding stable estimates of link closure and service-reduction probabilities. Sensitivity tests confirmed that increasing to 20,000 runs has minimal effect on link-level results. These probabilities provide one important measure of mobility over the often-long distances of links. To provide further insights to the expected magnitude of disruption and the required time and resources to repair and reopen roads, we also compute the number of road segments per link expected to: (i) be closed, $E[No.of\ closures] = \sum_i c_i$; and (ii) have reduced service, $E[No.of\ service\ reductions] = \sum_i r_i$, where $i$, $c_i$, and $r_i$ are as previously defined.

**Predicting Local Mobility in Two Counties of Focus: Pacific and Grays Harbor, WA**

To illustrate how road disruptions could affect post-event accessibility, we focus on two coastal counties where conditions converge to heighten isolation risk. Using the segment-level closure probabilities, we employ a Monte Carlo simulation to estimate the probability that each origin node (i.e., intersection) within the local road network can reach: (i) at least one of four hospitals in the two counties; and (ii) at least one of seven exits to other counties following a CSZ M9 rupture. For each simulation, road segments are randomly sampled as open or closed according to their modeled probabilities, while maintaining the spatial correlation structure derived from the fitted semivariogram model. Network connectivity is then evaluated using a shortest-path search algorithm, which determines whether a continuous traversable route exists between each origin and a destination. This process yields spatially distributed probabilities of "islanding" from key destinations and provides one measure of functional isolation, illustrating where emergency response and healthcare access may be most impacted.

**RESULTS AND DISCUSSION**

All outputs are digitally available from Sanger et al. (2026) and are here discussed in the order of: (i) segment-level outcomes, including demographic correlates and bridge impacts; (ii) link-level outcomes; and (iii) focused analyses in the two WA counties of special interest.

**Segment-Level Results**

Shown in Fig. 4(a) is a demonstrative example of the segment-level closure probabilities ($c_i$) for southwestern WA (note that link-level $c_i$ are shown in Fig. 4(b) and will be discussed later). Of the ~509,000 NHS segments in WA, OR, and CA, 976 (or ~88 km) have a greater than 15% $c_i$, most of which are in WA (Table 1). A total of 4,320 segments (or ~389 km) have more than 50% probability of service reduction ($r_i$), the majority of which are also in WA (Table 1). Summing probabilities across all segments, we compute the expected number of closed and reduced-service segments. For example, if ten segments each have $c_i$ = 0.10, the expected number of closures is one segment (i.e., $E[No.of closures] = \sum_i c_i$). By this estimation, we expect 2,999 closed segments, which equates to ~270 km, or 0.6% of NHS segments; analogously, we expect 19,168 reduced-service segments (~1,725 km, or 3.8% of NHS).

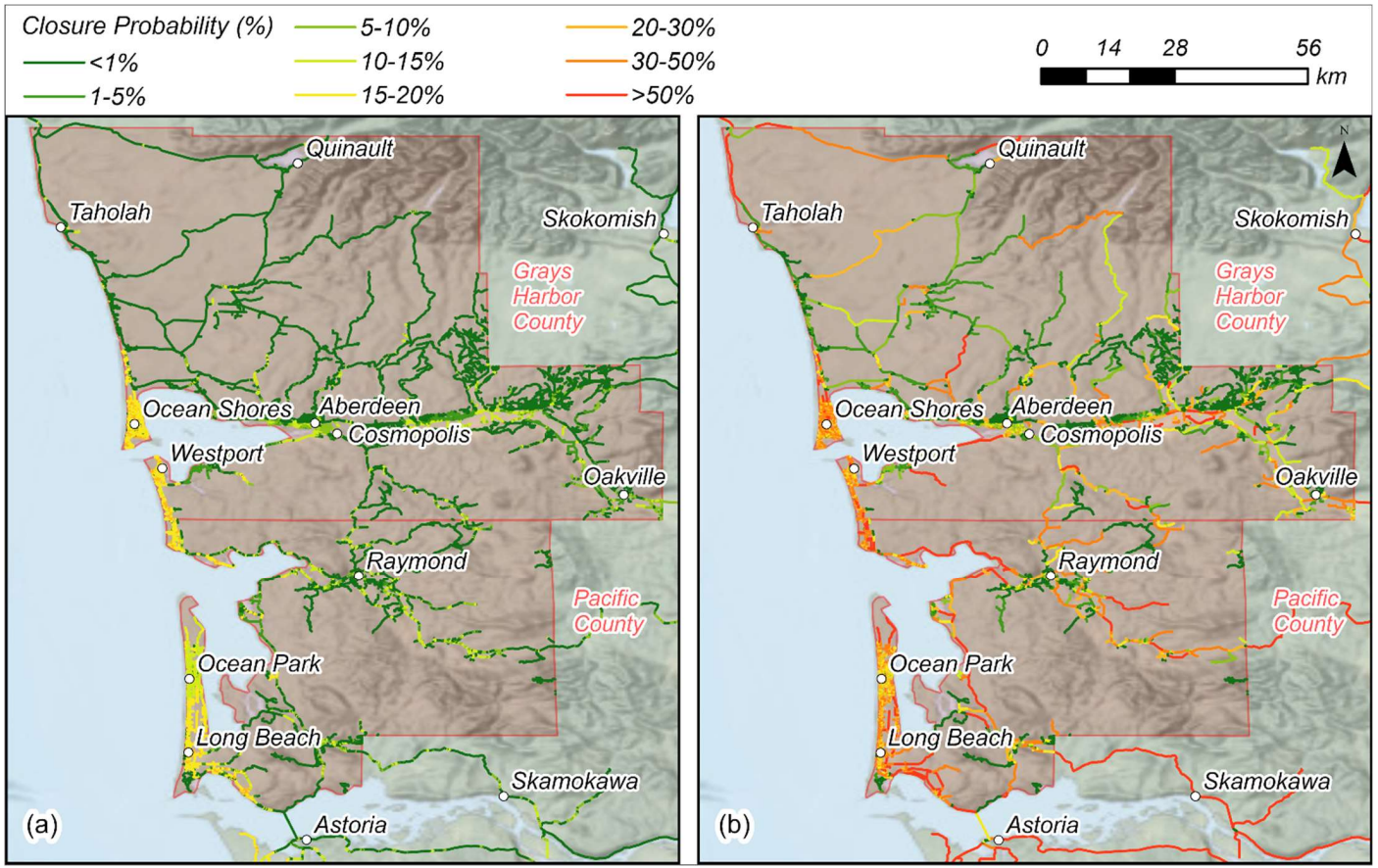

**Figure 4.** Probability of road closure in southwestern WA: (a) road segments; and (b) road links. All roads are shown within the two counties of interest, whereas only NHS roads are shown outside these counties.

The three counties most affected are Pacific (WA), Clatsop (OR), and Grays Harbor (WA), where the percentages of NHS segments with $c_i > 15\%$ are 7%, 4%, and 3%, respectively, and the percentages of NHS segments with $r_i > 50\%$ are 17%, 13%, and 6%, respectively. The NHS route most affected is U.S. 101, which hugs the coast from WA to CA: 228 segments (~21 km) have $c_i > 15\%$ and 3,985 segments (~359 km) have $r_i > 50\%$. Moving to the two counties of focus – Pacific and Grays Harbor – and considering all roads (i.e., not just NHS), 1,887 segments (~170 km, or 12% of roads) have $c_i$ > 15% and 2,610 segments (~235 km, or 17% of roads) have $r_i$ > 50%.

Table 1. Summary of predicted segment-level impacts for NHS roads.

| State | Probability of closure, $c_i$ > 15% | | Probability of service reduction, $r_i$ > 50% | |
| --- | --- | --- | --- | --- |
| | Number of segments | Equivalent length (km) | Number of segments | Equivalent length (km) |
| CA | 0 | 0 | 474 | 43 |
| OR | 336 | 30 | 1,806 | 163 |
| WA | 640 | 58 | 2,040 | 184 |
| **Total** | **976** | **88** | **4,320** | **389** |

**Road Impacts and Demographic Correlates**

To explore whether simulated road closures stratify socioeconomically, we overlaid 156 demographic indices (Esri, 2024) available at the resolution of census tracts and performed a statistical analysis stratified on *PGA*, enabling direct comparison among locations subject to similar shaking levels. To minimize bias from non-informative values, we limit these analyses to areas where shaking could conceivably trigger liquefaction, sampling all road segments with *PGA* > 0.1 g (N = 110,675). For each segment, measures of wealth, disposable income, home value, educational attainment, age, employment status, etc. were sampled, and segments were grouped into quantile-based bins of similar *PGA*. Within each bin the Pearson correlation coefficient between closure probability and each demographic variable was computed. These within-bin correlations were then pooled using a Fisher *z*-transformation weighted by bin sample size. The resulting coefficient, denoted $r_{PGA\text{-bin}}$, represents the association after controlling for *PGA* by stratification. Given the large sample size, most correlations – even if modest in absolute terms – were highly statistically significant ($p < 0.001$). Some of the strongest associations were observed with wealth and education metrics, as summarized in Table 2 (results for all 156 metrics are in Table S1 of the *Supplemental Materials*).

Groups with higher measures of wealth, home ownership, home value, education, and age are negatively correlated with road closure probabilities, suggesting that wealthier, older, and more educated homeowner communities may be less likely to face severe disruptions. Conversely, positive correlations are observed in populations with lower income and less educational attainment, indicating greater exposure.

Greater daytime worker population also showed positive correlation, indicating that low-laying hubs of industry and commerce may be impacted more.

Road closures in CSZ scenarios have the potential to disrupt access to critical services at a regional scale, and systematic, albeit small, demographic gradients in exposure may amplify pre-existing economic inequalities. All demographic correlations are "weak" per standard benchmarks, but at the population level can amount to disproportionate burdens on large numbers of people; these correlations are similar in magnitude, for example, to those measured in economic stratifications of air pollution (Collins et al., 2022). These results provide an initial indication that socioeconomic vulnerability may compound with physical hazard exposure. Further analysis is warranted to elucidate and quantify these trends.

Table 2. Select demographic correlates with closure probability, ranked by Pearson's coefficient.

| Rank | Metric | $r_{PGA\text{-bin}}$ | $p$-value |
|---|---|---|---|
| 1 | Economic Dependency Ratio | -0.087 | < 0.001 |
| 2 | Homeownership rate | -0.083 | < 0.001 |
| 3 | Home Value $1,000,000-$1,499,999 | -0.075 | < 0.001 |
| 4 | Wealth Index | -0.062 | < 0.001 |
| 5 | Population Age 25+: Graduate/Professional Degree | -0.048 | < 0.001 |
| 1 | Population Age 25+: Less than 9th Grade Education | +0.157 | < 0.001 |
| 2 | Daytime Population: Workers | +0.147 | < 0.001 |
| 3 | 2024 Renter Occupied Housing Units | +0.091 | < 0.001 |
| 4 | Household Income $15,000-$24,999 | +0.083 | < 0.001 |
| 5 | Home Value less than $50,000 | +0.067 | < 0.001 |

**Analysis of *DHS (2019) Bridges***

DHS (2019) assumed any bridge on a soil profile predicted to liquefy to any degree would suffer "significant" damage and closure. This resulted in the expectation of 2,669 closed bridges in western WA following an M9 CSZ event, as mapped in Fig. 5(a). Although a comprehensive simulation would require site-specific bridge and soil data coupled with more advanced modeling, we here use the segment-level results to preliminarily analyze the 6,911 bridges studied by DHS (2019). Although bridge foundations are not explicitly modeled, the segment-level predictions of liquefaction manifestation severity may reasonably approximate damage to bridge approaches and embankments and serve as a proxy of overall bridge disruption. The adopted approach is thus simplifying, like DHS (2019), but is – in the authors' opinion – supported by more rigorous models and more consistent with global observations of bridge functionality following strong shaking and/or liquefaction (e.g., Palermo et al., 2011, 2017; Lew et al., 2021).

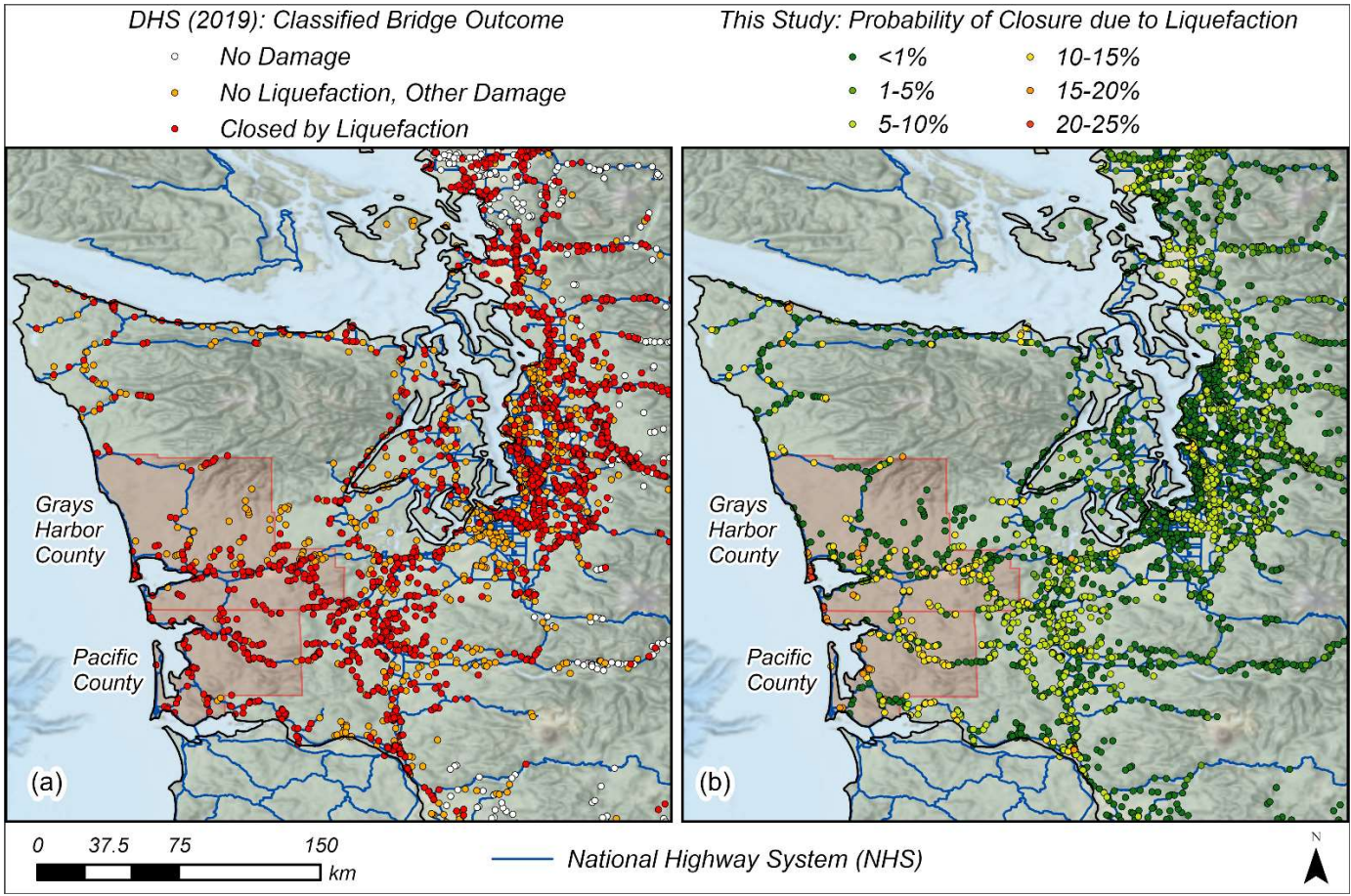

**Figure 5.** Predicted impacts to 6,911 WA bridges in an M9 CSZ rupture: (a) DHS (2019); (b) this study.

DHS (2019) predicted that 316 of the closed bridges would experience no source of damage other than liquefaction. Shown in Fig. 6(a) are the cumulative distributions of these 316 bridges as a function of outcomes predicted in this study. The median probabilities of closure, service reduction, and any surface manifestation of liquefaction are 1%, 10%, and 18%, respectively. For 38% of these bridges, the probability of manifestation is zero, suggesting that lateral spreading hazard – which is not explicitly modeled – would also be negligible. None of the 316 bridges have a probability of closure exceeding 9% or a probability of service reduction exceeding 42%. Summing probabilities across all 316 bridges, we expect 61 bridges with *some* surficial expression of liquefaction, 34 with reduced service (e.g., speed reduction), and 7 closures.

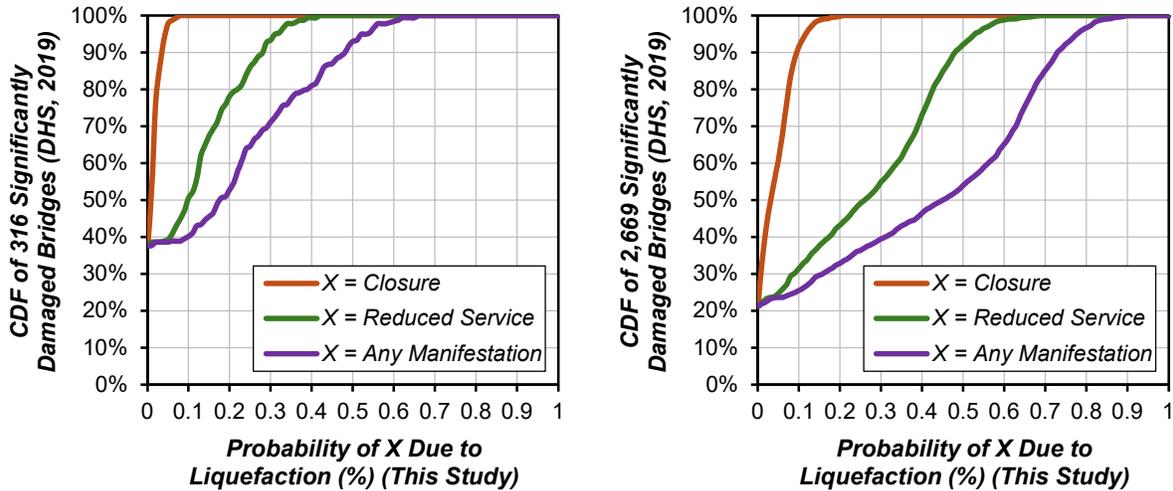

**Figure 6.** Cumulative frequency of bridges with "significant damage" due to liquefaction per DHS (2019) as a function of predicted outcomes in this study: (a) 316 bridges where liquefaction was the only cause of damage; (b) 2,669 bridges where liquefaction was not the only cause of damage.

DHS (2019) also predicted a total of 2,669 bridge closures where liquefaction was sufficient for closure but was not necessarily the only cause of damage. Cumulative distributions are plotted in Fig. 6(b) for these bridges and portray a similar outcome. None of the 2,669 bridges have a closure probability exceeding 21% or a service-reduction probability exceeding 68%. For 21% of these bridges, the probability of liquefaction manifestation is zero. Summing probabilities across all 2,669 bridges, we expect 1,048 with *some* surface expression of liquefaction, 655 with reduced service, and 110 closures. Thus, of the 2,669 closed bridges expected by DHS (2019), we predict just 4% will be closed (of course, this could still obstruct the welfare, economy, and recovery of the region). It is also worth noting that recent earthquakes elsewhere provide evidence of bridges functioning even after large liquefaction-induced deformations (Palermo et al., 2011, 2017), albeit any evaluation of this possibility would require bridge- and foundation-specific modeling of different failure modes. Although our analysis does not consider all mechanisms of damage, we believe it to be a more accurate approximation of liquefaction impacts on the portfolio of bridges in western WA. The results mapped in Fig 5(b), and digitally available, provide a starting point from which bridges may be prioritized for more advanced analysis and, if ultimately appropriate, hazard mitigation and retrofitting.

**Link-Level Results**

Shown in Fig. 4(b) is a demonstrative example of the link-level $c_i$ in southwestern WA. Contrasting with Fig. 4(a), it is seen that relatively low segment-level $c_i$ can give rise to high probabilities of link closure. This is because links may contain many hundreds or thousands of segments and just one closure is needed to close the link. Link performance thus reflects both link length and the performance of individual segments. Of the ~12,000 NHS links across WA, OR, and CA, 1,586 have greater than 15% $c_i$ and 2,070 have greater than 50% $r_i$, most of which are in OR (Table 3). Summing probabilities across all links, we expect a total of 745 closed links (6% of NHS links in study area) and 2,096 reduced-service links (17%). Counties most affected include Pacific (WA), Tillamook (OR), and Grays Harbor (WA), where the percentages of NHS links with $c_i$ > 15% are 91%, 79%, and 76%, respectively, and the percentages of NHS links with $r_i$ > 50% are 91%, 93%, and 90%, respectively. Notably, the counties with the greatest *quantity* of impacted links are those most urbanized, but because such counties have large, redundant networks, the percentage of links impacted is less. King County (WA), for example, has 180 NHS links with closure probability > 15%, which represents 18% of NHS links in the county. The NHS routes with greatest expected impact include: (i) U.S. 101: 78 links (33%) have $c_i$ > 15% and 85 links (36%) have $r_i$ > 50%; and (ii) Interstate 5: 115 links (31%) have $c_i$ > 15% and 144 links (39%) have $r_i$ > 50%. Within the two counties of focus – Pacific and Grays Harbor – and considering all roads, 4,483 links (39%) have $c_i$ > 15% and 5,701 links (49%) have $r_i$ > 50%.

**Table 3**. Summary of predicted link-level impacts for NHS roads.

| State | Probability of closure, $c_i$ > 15% | | Probability of service reduction, $r_i$ > 50% | |
| --- | --- | --- | --- | --- |
| | Number of links | Proportion (%) | Number of links | Proportion (%) |
| CA | 164 | 10 | 187 | 9 |
| OR | 825 | 52 | 1,103 | 53 |
| WA | 597 | 38 | 780 | 38 |
| **Total** | **1,586** | -- | **2,070** | -- |

The preceding probabilities provide one important measure of link performance, but do not explicitly quantify the number of segment failures within a link. A link could have high $c_i$ because a single segment has very high $c_i$, or because many segments have small but non-zero $c_i$, for example. To evaluate the expected magnitude of disruption and the required time and resources to repair and reopen links, we compute the number of segments per link expected to be closed and have reduced service, as demonstrated in Fig. 7 for southwest WA. As previously stated, we expect a total of 2,999 closed segments (~270 km) and 19,168 segments (~1725 km) with reduced service. The NHS route with the greatest number of expected segment impacts is U.S. 101, with 284 expected closures and 1,495 expected service reductions. The counties most impacted by NHS closures in terms of sheer quantity include counties in the vicinity of Portland (OR) and Seattle (WA), whereas those with the highest percentage of closed NHS segments include a mix of urban and coastal counties.

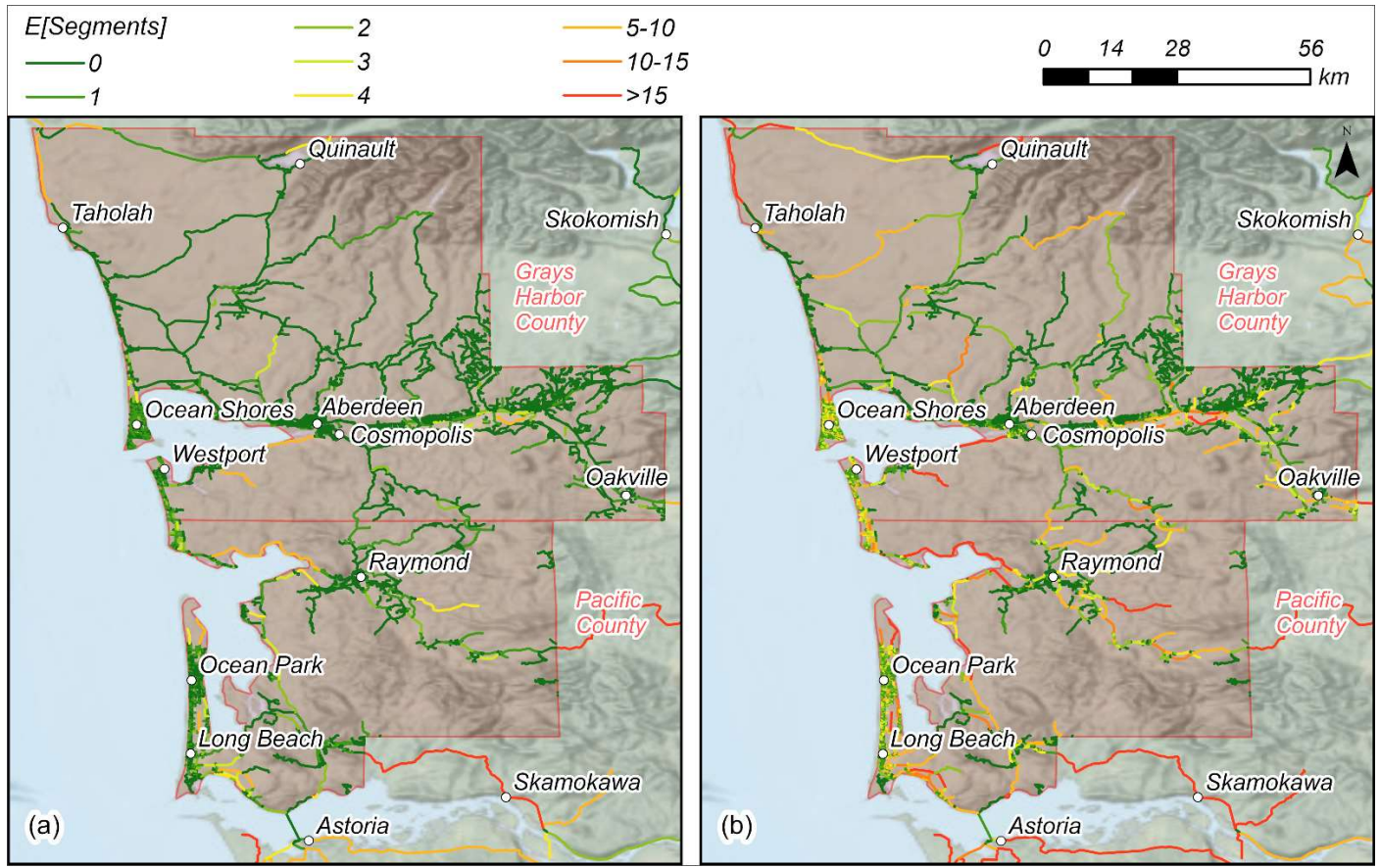

**Figure 7.** Expected number of segments per link: (a) closed; and (b) with reduced service. All roads are shown within the two counties of interest, whereas only NHS roads are shown outside these counties.

Considering all roads within the two counties of focus highlighted in Fig. 7, a total of 2,131 segments (~191 km, or 18.4% of county segments) are expected to be closed, and 8,993 segments (~809 km, or 77.5% of county segments) are expected to have service reductions. What these analyses do not explicitly account for, however, is network redundancy and broader mobility. In other words, can the links expected to be closed be circumnavigated, or will these closures hinder emergency response, recovery, commerce, etc.? As seen in Fig. 7, many towns in these coastal counties have limited network redundancy, with just 1 or 2 links connecting elsewhere, and in some cases these "lifeline" links are expected to have multiple closures.

**Mobility in Pacific and Grays Harbor Counties**

Monte Carlo simulations of network connectivity provide a preliminary evaluation of post-event mobility in the two counties of focus. Fig. 8 summarizes node-based accessibility across the regional road network, expressed as the probability of reaching: (a) at least one of four hospitals; and (b) at least one of seven exits to NHS roads. Probabilities are computed for each node, capturing the combined effects of network topology, spatial distribution of destinations, and probabilistic link disruption. Fig. 8(a) shows that hospital access is generally highest within and immediately surrounding the more urban, inland centers such as Aberdeen, Hoquiam, and Cosmopolis, where hospitals are relatively close, the road network is denser and more redundant, and the modeled liquefaction hazard – while often high – is lower than on the immediate coast. In these areas, most nodes have 90% connection probabilities. In contrast, markedly lower probabilities are computed along the coastal margins of the region (e.g., Ocean Shores, Westport, Long Beach, and Ocean Park) where access probabilities frequently fall below 2%, reflecting longer travel distances to hospitals, limited network redundancy, and higher likelihoods of damaging liquefaction manifestations. These zones are characterized by barrier spits (i.e., peninsulas) and corridors with few alternative routes. Closure probabilities for individual road segments remain modest on these peninsulas (e.g., < 10%), as seen in Fig. 4(a), but access to the mainland requires numerous segments in series to remain functional, and sharp accessibility gradients occur where networks bottleneck through critical links.

Compared to hospital access, Fig. 8(b) shows that egress via NHS roads is generally expected to be similarly affected. Nodes near major highway corridors exhibit high probabilities, often exceeding 90%. Yet large portions of the coastal network again show very low probabilities of access. Notably, some nodes with high hospital-access probabilities display substantially lower NHS-access probabilities, indicating that local mobility may remain feasible even when access to long-distance evacuation routes is unlikely. In some communities (e.g., Raymond), it is apparent that hospital access benefits from facilities embedded within urban networks, whereas NHS access is concentrated along a limited set of corridors. As a result, the loss of a small number of links can disproportionately reduce evacuation potential without equivalently degrading access to nearby medical care.

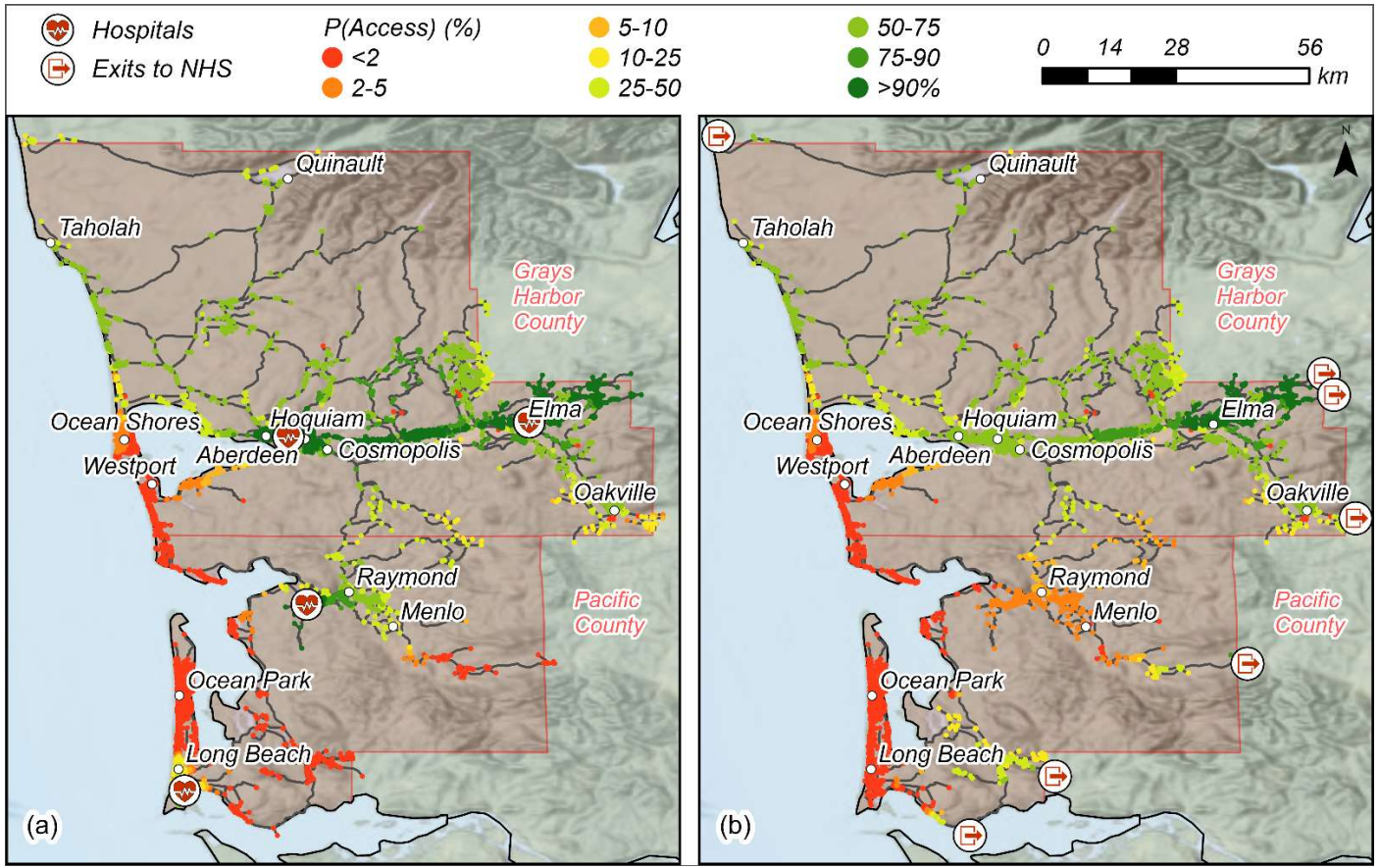

**Figure 8.** Probability of travelling by road network to at least one: (a) hospital; and (b) exit point to NHS.

Summarily, these preliminary results suggest that communities with low probabilities in both panels, particularly coastal and peninsular areas, face compounded isolation risk. Identifying nodes and corridors that control transitions between high- and low-probability regions provides a basis for targeted, site-specific sampling and detailed analysis. In interpreting these results, it should be noted that the longevity of closure is not considered, nor is the potential for alternate routing or emergency repairs, the feasibility of which may depend on a road's topographic positioning and the redundancy of the surrounding network (Daglish et al., 2025). Also, the possibilities of reaching hospitals outside the two counties, or exiting the counties by other means (e.g., ports, airfields) are not considered. Most importantly, the very high liquefaction hazards modeled by the GLM across the barrier peninsulas and coastal estuaries in Grays Harbor and Pacific counties have minimal ground truthing. Of the 37,000 CPTs used to train and locally update the ML model, just 21 are sited within these counties and more broadly, there are relatively few CPTs from similar geomorphological settings, given the typically rural nature of coastal communities in the U.S. Pacific Northwest. It is thus imperative that more geotechnical data be to confirm or modify the ML predictions.

**Limitations and Uncertainties**

This study is subject to limitations that arise from uncertainties in earthquake hazard modeling, liquefaction prediction, infrastructure fragility representation, and network analysis assumptions. These limitations are important for interpreting and contextualizing the results for planning, policy, and risk communication.

*First*, ground motion inputs are based on the median predictions from physics-based simulations of a CSZ M9 rupture. While this approach improved realism relative to single-scenario analyses, it does not capture the full range of plausible rupture characteristics. Local deviations in shaking intensity could be higher or lower than modeled, especially in sedimentary basins and near rupture asperities. Additionally, the median representation does not explicitly quantify uncertainty bounds, which may be important for probabilistic risk management but remain outside the scope of this study.

*Second*, liquefaction hazard is modeled using a geospatial ML framework trained on CPT data. This method enables continuous spatial coverage, but the predictions are constrained by the resolution and quality of regional datasets, such as groundwater depth, geomorphologic indicators, and mapped surficial geology. Errors can arise where the true subsurface conditions differ materially from those suggested by regional proxies or where fine-scale heterogeneity is uncaptured by geospatial variables. Like all data-driven methods, the model reflects correlations present in the training data, which do not always imply physical causation. Although the framework is informed by mechanics and employs safeguards against overfitting, performance will continue to improve as more in-situ data become available, particularly in regions underrepresented in training. Another source of uncertainty concerns sampling bias in subsurface data. CPTs tend to be collected preferentially in locations where liquefaction is expected or where field conditions permit testing, which may skew results toward susceptible settings. While prior validation efforts show no strong evidence of bias (Sanger et al., 2025), regions lacking CPT data lack ground truthing. Integration of complementary datasets, including standard penetration test data and borehole logs, should improve model robustness in such environments (e.g., as discussed in Grays Harbor and Pacific counties).

*Third*, the translation from expected liquefaction manifestation to road impacts is necessarily approximate, based on case histories of liquefaction surface expression, empirical evidence of liquefaction impacts to roads, and analogs from flooding, snow accumulation, and volcanic ash impacts. Road functionality in practice likely depends on additional factors, including maintenance standards, pavement condition, traffic control infrastructure, emergency management decisions, and post-event debris clearance. These translations also do not explicitly model lateral spreading, which remains one of the most damaging forms of liquefaction and often occurs preferentially near free faces or on sloped terrain. Therefore, closure probabilities may be underestimated in river valleys, coastal settings, and reclaimed waterfronts where lateral spreading is more likely. The assignment of non-zero probability of closure at lower *LPI* values aims to account for this limitation but does not fully capture the underlying mechanical controls. High-resolution elevation and bathymetry modeling could support additional analyses of lateral spreading in the future.

*Fourth*, the analyses do not explicitly represent roadway design features, such as pavement thickness, embankment geometry, drainage systems, or ground improvement. All roads are implicitly treated as structurally similar and equally vulnerable to equivalent levels of surface deformation. Improved ground or pavement systems that are newer and more robust than average, may perform better, while older and less robust roads may perform worse. Bridge impacts are treated in a similarly simplified manner, using liquefaction predictions at the ground surface as a proxy for damage to approaches and foundations. This approximation does not consider foundation type or bridge-specific features, which have been compiled for some bridges in western WA (Liu et al., 2022) and could be further studied. *Fifth*, network connectivity analyses assume binary states of closure or service reduction and do not consider traffic, congestion, emergency rerouting, or duration of impacts. The results therefore describe the probability of functional disconnection, not expected changes to travel time, travel distance, network efficiency, or network recovery following the event. *Additionally*, the socioeconomic analyses use survey data that is subject to measurement error, and which is available at the scale of census tracts, which have a median area of 4.6 km$^2$ (1.8 miles$^2$) and are generally much smaller in urban areas. The results should not be interpreted as applicable to individual households. Moreover, these analyses considered simulated road-segment closures, which do not necessarily equate to reduced mobility or reduced access to critical facilities. The analyses should thus be considered preliminary and deserving of further analysis.

*Finally*, this work is not intended to replace site-specific geotechnical investigations or detailed transportation resilience studies. Rather, it provides a standardized regional assessment framework designed to reveal systematic vulnerabilities and inform prioritization at the regional scale. Local studies, incorporating detailed site characterization and design data, remain essential for decision-making related to retrofitting, operations planning, and investment. Studies of other earthquake impacts on road networks (e.g., tsunamis, landslides, surface faulting, bridge superstructure damage) are also important. Despite these limitations, this study represents the first systematic effort to quantify liquefaction-driven disruption to road networks across the CSZ using mechanics-informed, data-driven models. As geotechnical databases expand and future earthquakes provide new observations, the methods and results presented here can be further refined and validated.

**CONCLUSIONS**

This study presents a mechanics-informed, data-driven framework for evaluating liquefaction-induced disruption of road networks in Cascadia Subduction Zone earthquakes. By integrating a geospatial liquefaction model with empirically derived fragility relationships and network analysis, we quantify the probability of closure and reduced service across transportation corridors in the U.S. Pacific Northwest. Results indicate that impacts are strongly concentrated in coastal zones, river valleys, and urban waterfronts, with particularly high isolation risk in Pacific and Grays Harbor Counties, Washington, where limited network redundancy and extremely high modeled hazard exacerbate loss of access to critical facilities. Compared with prior statewide assessments, our findings suggest substantially lower numbers of liquefaction-driven bridge closures, reflecting improved representation of subsurface conditions and damage mechanisms. While large uncertainties remain, the results provide a regional baseline for emergency planning, asset screening, and prioritization of more advanced analysis and, potentially, mitigation. As geotechnical datasets expand, this framework offers a pathway toward increasingly accurate regional assessments of infrastructure vulnerability to liquefaction.

## DATA AVAILABILITY STATEMENT

Some or all data, models, or code generated during this study are available from the corresponding author. Additionally, many digital GIS products are provided in the supplemental materials described below.

## ACKNOWLEDGMENTS

This work was facilitated using high performance computing infrastructure provided by the Hyak supercomputer funded by the University of Washington's student technology fund, and through DesignSafe at the Texas Advanced Computing Center (Rathje et al. 2017).

## SUPPLEMENTAL MATERIALS

Eq. S1, Table S1, and many GIS files are available from Sanger et al. (2026), including all salient products appearing in Figs. 1, 2, 4, 5, 7, and 8 of the main text.

## FUNDING

The presented work is based on research supported by the United States Geological Survey (USGS) under award G23AP00017, the Cascadia Region Earthquake Science Center (CRESCENT) via National Science Foundation (NSF) award 2225286, the Cascadia Coastlines and Peoples (CoPes) Hub via NSF award 2103713, the Pacific Earthquake Engineering Research (PEER) Center under award 1185-NCTRMB, and the Pacific Northwest Transportation Consortium (PacTrans) under award 69A3552348310. However, any opinions, findings, conclusions, or recommendations expressed herein are those of the authors and may not reflect the views of USGS, NSF, CoPes, CRESCENT, PEER, or PacTrans.